\documentclass[fleqn,usenatbib]{mnras}

\usepackage[T1]{fontenc}
\usepackage{ae,aecompl}

\usepackage{graphicx}	
\usepackage{amsmath}	
\usepackage{amssymb}	
\usepackage{tabularx,booktabs,tablefootnote}
\usepackage{multirow}
\usepackage{hyperref}
\usepackage{newtxtext,newtxmath}
\usepackage{multicol}
\usepackage{subcaption}

\title[Tayler instability in 56 Ari]{Tayler instability as possible reason for the period changes in Ap star 56 Ari.}

\author[I. Potravnov \& L. Kitchatinov]{
I.S. Potravnov$^{1,2}$\thanks{E-mails: ilya.astro@gmail.com (IP), kit@iszf.irk.ru (LK)},
L.L. Kitchatinov$^{2}$\footnotemark[1],
\\
$^{1}$Institute of Astronomy of RAS, Pyatnitskaya str., 48, 119017, Moscow, Russia\\
$^{2}$Institute of Solar-Terrestrial Physics Siberian Branch of RAS, Lermontova str., 126A, 664033, Irkutsk, Russia \\}

\date{}

\pubyear{2025}

\begin{document}
\label{firstpage}
\pagerange{\pageref{firstpage}--\pageref{lastpage}}
\maketitle

\begin{abstract}
The physical mechanism responsible for the photometric period changes in chemically peculiar star 56 Ari was searched. It was previously shown that rate of the star's period increase is few orders of magnitude larger than the rates expected from the evolutionary changes of the angular momentum or due to magnetic braking. Also no secular changes were detected in the surface structure or visibility of chemical spots which are responsible for the rotational modulation of stellar brightness. We hypothesise that period changes in 56 Ari are caused by the drift of surface magnetic and associated abundance structures as a result of the kink-type (Tayler) instability of the background magnetic field in the radiative zone of the star. Results of the numerical simulation presented in the paper yield growth and drift rates of the most rapidly developing non-axisymmetric mode of the instability, consistent with the observed rate of period changes in 56 Ari. The surface geometry of the 56 Ari magnetic field is also reproduces in the calculations. The proposed mechanism may also be used to explain the character of period changes in other Ap/Bp stars demonstrating such an effect.
\end{abstract}




\section{Introduction}

One of the distinctive observational features of chemically peculiar magnetic Ap/Bp stars is their synchronous photometric, spectral and magnetic variability. Such an variability can be explained within framework of the so-called "oblique rotator" model\, \citep{Stibbs_1950} by the axial rotation of the chemically inhomogeneous (spotty) stellar surface with magnetic axis inclined to the line of sight. The surface chemical composition of Ap/Bp stars is built in by the selective atomic diffusion in magnetized atmosphere \citep{Michaud_1970,Michaud_2015}. The magnetic field controls the ions transport and their surface distribution follows local geometry of the field \citep{Alecian_1981,Michaud_1981}. As a consequence, the horizontal distribution of elements in the atmospheres of Ap/Bp stars has the character of chemical spots with altered thermal structure due to modified opacity and lines blanketing. The coincidence of magnetic, photometric and spectral periods of Ap/Bp stars confirms the relation between chemical spots and magnetic structures. 

Within the framework of the currently prevailing hypothesis of the relic origin of the large-scale magnetic fields of Ap/Bp stars \citep{Cowling_1945,Moss_2001}, the fields after their generation are considered stable in the radiative envelopes at timescale of $\sim10^8-10^9$ yr, comparable to the lifetime on the Main Sequence (MS). Therefore, after rapidly ($\sim10^4$ yr \citep{Alecian_2010,Stift_2016}) shaping an equilibrium abundances distribution by the vertical diffusion, one can expect stability of chemical spots and photometric periods during the evolution of the star on the MS. Indeed, the periods of most well-studied Ap/Bp stars are preserved with high precision throughout their observational history. A canonical example is the $\alpha^2$ CVn, which period determination by \citet{Belopolsky_1913} is still valid \citep{Catalano_1998}. 

However, there is a small subgroup of Ap/Bp stars contains about dozen of objects which display photometric phase shifts, potentially indicative for the changes in their periods \citep[see][for review]{Pyper_2020,Pyper_2021}. This is a fine effect with a typical rate of $\dot P\sim10^{-8}$ d/cycle. Nevertheless, it is confidently detected on long-term series of high-precision photometric observations. The direction of change can vary: the phase shifts observed for most objects are best fitted with linear period increase model. Nevertheless, few objects show period decreases, and in the unique case of the ApSi star CU Vir the phase shifts exhibit discrete glitches detected from photometric and spectroscopic data \citep{Pyper_2020,Pyper_1998,Pyper_2013}, as well as from radio observations \citep{Trigilio_2008}. An alternative interpretation of the observations for some objects: CU Vir, V901 Ori, V913 Sco are quasi-sinusoidal period changes \citep{Mikulasek_2011,Mikulasek_2016,Shultz_2019}, although for V901 Ori and V913 Sco a reversal of the sign of the period changes has not been confirmed \citep{Pyper_2020}. In the case of CU Vir, the O-C diagrams for the model with discrete and continual period changes are comparable in accuracy \citep{Pyper_2020}.

Given such a diverse character of period changes and their rates, it is difficult to explain them by the evolution of the angular momentum of the star as a whole \citep[e.g.][]{Stepien_1998,Potravnov_2024}. For CU Vir there is a fundamental contradiction between the braking time and the age of the star \citep{Mikulasek_2011}. As alternative explanations for the observed photometric phase shifts, variations in (a) the surface distribution of the spots \citep{Adelman_1992,Pyper_1998} or (b) their visibility due to the free-body precession of the rotation axis of the magnetically-distorted star \citep{Shore_1976} were discussed.

For the case of the silicon star 56 Ari, one of the first Ap/Bp stars with detected photometric minimum phase shifts \citep{,Musielok_1988,Adelman_2001,Pyper_2021}, this issue was investigated in \citep{Potravnov_2024}. With the Doppler imaging technique the maps of surface distribution of silicon were recovered on an interval of $\approx30$ yr, comparable to the interval on which phase shifts were detected. Silicon spots are the key contributors to the light modulation in Ap stars in the considered temperature range \citep[e.g.][]{Krtichka_2012,Pakhomov_2024}. The recovered maps revealed the stability of the spots configuration over the discussed time interval and, within the accuracy of the method, the absence of latitudinal shifts of the spots due to precession. Thus, the spots and magnetic structures in 56 Ari rotate rigidly, and the observed phase shifts indicate a linear increase in the rotation period with rate of $\Delta P\approx2-4$ s/100 yr \citep{Adelman_2001,Pyper_2021}. This rate is $\sim2$ orders of magnitude larger than the expected rate of angular momentum loss during the evolution of the star. The question of the reasons for the period changes in 56 Ari and other Ap/Bp stars of the sample remains unresolved.

An alternative group of mechanisms which can affect the observed periods of Ap/Bp stars are the "magneto-rotational"\, effects producing azimuthal variations of the magnetic field and the accompanying drift of the chemical structures. Thus, \citet{Mestel_1987} consider torsional Alfvén waves arising due to the disturbance of the isorotation of the magnetised atmosphere and the inner layers of the star. In \citet{Stepien_1998,Krtichka_2017} this mechanism was used to explain period variations in CU Vir, assuming their continual character (although, as noted above, alternative interpretations of the observations of this star are also possible). In a recent paper by \citet{Takahashi_2024}, standing Alfvén waves were analysed using models of the stellar internal structure and it was shown that their most probable oscillation frequencies can potentially explain the period variations in Ap/Bp stars and the effect should be periodic. 

Another possibility to explain the periods changes in the Ap/Bp stars within this group of mechanisms is the so-called Tayler instability. In the original papers by \citet{Tayler_1973,Pitts_1985} it was shown that the axisymmetric toroidal field in the radiative zone of a star is unstable to small non-axisymmetric perturbations. In subsequent works \citep{Spruit_2002,Rudiger_2012} it was found that this instability, which came to be referred as the Tayler instability, may be responsible for the amplification and the observed surface configuration of the magnetic fields of Ap/Bp stars. An important feature of the Tayler instability is the drift of its azimuthally inhomogeneous modes against the direction of the stellar rotation \citep{Rudiger_2010,Kitchatinov_2020}. In a reference frame rotating with the star, the instability pattern drifts in the counter-rotational direction without changing its spatial structure. This "rigid"\, drift of surface magnetic structures and associated chemical spots can affect the observed periods of Ap/Bp stars. In the present paper, we use MHD modelling for investigation the possibility that the increase in the photometric period of the ApSi star 56 Ari is caused by the Tayler instability of its magnetic field.

\section{Ap star 56 Ari and its period changes}

56 Ari (=HD 19832) was classified as peculiar A0 star with enhanced lines of ionised silicon \ion{Si}{II} 4128/4131 \AA\, in the HD catalogue \citep{Cannon_1918}. A modern accurate determination of the atmospheric parameters and chemical composition of 56 Ari was performed in \citep{Shulyak_2010}. The effective temperature of the star $T_{eff}=12800$ K, as well as an excess of silicon and iron in its atmosphere together with a deficit of light elements: helium and magnesium was found. In general, the parameters of 56 Ari are typical for chemically peculiar Ap/Bp stars. Summary of the parameters of 56 Ari is given in the Table \ref{tab1} with the corresponding references. An important feature of 56 Ari is the atypically large for Ap/Bp stars projected axial rotation velocity $V\sin i\approx$165~km/s \citep{Hatzes_1993,Potravnov_2024}. Combination this value with the stellar radius obtained from the observed spectral energy distribution, $R/R_{\odot}=2.38$, and the photometric period (see below) results in the angle of inclination of the stellar rotation axis to the line of sight $i\approx80^\circ$. This means that the star is seen almost equator-on. The results of Doppler imaging \citep{Hatzes_1993,Ryabchikova_2003,Potravnov_2024} reveal a inhomogeneous surface distribution of He, Mg, Si, and Fe arranged in chains of near-equatorial spots. The magnetic field of 56 Ari was investigated with spectropolarimetry \citep{Borra_1980,Shultz_2020}. The amplitude of variations of the longitudinal field <$B_Z$> was about 800 G, which corresponds to the dipole component strength $B_d\approx2.7$ kG. Fitting of the magnetic curve also results in a large inclination of the magnetic axis to the rotational one $\beta\approx80-90^\circ$. Thus, the magnetic poles lie in the plane of the rotational equator. Comparison of the magnetic curve with surface maps of silicon distribution in 56 Ari allows to deduce qualitatively that the magnetic poles correspond to regions of reduced silicon abundance \citep{Potravnov_2024}.    

\begin{table}
\caption{Parameters of 56~Ari}
\label{tab1}
\begin{tabular}{lcc}
\hline
Parameter & Value & Reference \\
\hline
\hline
$T_{eff}$ & 12800$\pm$300 K & 1 \\
$\log g$  & 4.0$\pm$0.05 dex & 1 \\
$V\sin i$   & 165$\pm$5~km/s & 2 \\
$i$      &$80^\circ\pm10$ & 2\\ 
$R/R_{\odot}$       & 2.38$\pm$0.2& 2\\
$\log (L/L_{\odot})$ & 2.14$\pm0.07$ & 2 \\
$M/M_{\odot}$       & 3.2$\pm0.3$ & 2,3 \\
$B_d$               & 2.7 kG & 4 \\
$\beta$               & 80-90$^\circ$ & 4,5 \\
$\log t$               & 7.95 y & 3 \\
\hline
\end{tabular}
\bigskip

\emph{Note} 1 - \citet{Shulyak_2010}; 2- \citet{Potravnov_2024}; 3 - \citet{Kochukhov_2006}; 4 - \citet{Shultz_2020}; 5 - \citet{Borra_1980} 
\end{table}  

In combination with inhomogeneous surface elemental distribution, the fast axial rotation leads to modulation of light and spectral lines profiles with a short period $P\approx0.728^d$ for the first time correctly established in \citet{Deutsch_1953,Provin_1953}. The light curve of 56 Ari exhibits two peaks at phases $\phi \approx 0.3$ and 0.7, corresponding to the passage of two groups of silicon spots through the central meridian. The shape of the visual light curve and the phases of its extrema correlate with the variation of the equivalent widths of the silicon \ion{Si}{II} 6347/6371 \AA\, lines \citep{Adelman_2001}. Because of the double-peaked shape of the maximum, the period of this star is determined from the minimum light.

A phase shifts of the 56 Ari photometric minimum corresponding to a period increase of 4 s/100 yr was recorded for the first time in \citet{Musielok_1988}, based on archival data and original observations during the interval 1952-1988. Based on an analysis of photometric and spectroscopic observations of 56 Ari from 1952 to 2001, in \citet{Adelman_2001} the authors confirmed the phase shift of the minimum, which in the O-C analysis is best described by a model of a linearly increasing period with a rate of 2 s/100 yr. The existence of a secondary period with a length of $\sim5$ yr, produced by free-body precession of the stellar rotation axis, was also suspected. Homogeneously obtained photometry with the FCAPT automatic telescopes in the four-colour Stromgren system has extended the series of observations of 56 Ari up to 2013 \citep{Pyper_2021}. Analysis of this long-term photometry revealed a somewhat larger rate of linear period change within the linear model: 4.11 s/100 yr, in agreement with an earlier estimation \citep{Musielok_1988}. \citet{Pyper_2021} also increased the lower limit on the possible precession period up to $\sim$30 yr. However, no differential changes in the structure of silicon spots and their visibility were detected over a comparable time interval \citep{Potravnov_2024}. In the present paper, we assume that the changes in the period of 56 Ari have the character of a linear increase with $\Delta P\simeq$2-4 s/100 yr.

\section{The model}

Calculations of the Tayler instability in 56 Ari were performed with the theoretical model developed in \citet{Kitchatinov_2008a,Kitchatinov_2008b}. The model equations and the method of their numerical solution are discussed in detail in these papers. Therefore, we will only review its basic assumptions and determine the values of the model parameters used in the calculations for 56\,Ari.  

As it was shown by \citet{Spruit_1999}, the dominant component of the magnetic fields in the radiative interiors of the stars is the axisymmetric toroidal field $B$. For the aims of the present work, it is convenient to define such a field in terms of the angular Alfven velocity $\Omega_A$,
\begin{equation}
   B = \sqrt{4\pi\rho}\sin\theta\,r\Omega_A\,,
   \label{1}
\end{equation}
where $\rho$ is density and the conventional spherical coordinates are used. The problem of stability of a rotating star with magnetic field ({\ref{1}) to small perturbations of magnetic field ($\vec{b}$) and velocity ($\vec{v}$) was solved. In spherical geometry, it is convenient to represent perturbations as a superposition of their toroidal and poloidal components \citep{Chandrasekhar_1961}. Then for perturbations of the magnetic field we obtain    

\begin{equation}
    \vec{b} = \vec{r}\times\mathrm{\nabla}(T/r) 
    + \mathrm{\nabla}\times(\vec{r}\times\mathrm{\nabla}(P/r)).
    \label{2}
\end{equation}
For this formulation, the force lines of the toroidal field lie on spherical surfaces of constant radius and are isolines of the toroidal potential $T$. Isolines of the radial component of the field $b_r$ coincide with the isolines of the function $P$ on the spherical surface. 

\begin{figure}
\includegraphics[width=\linewidth]{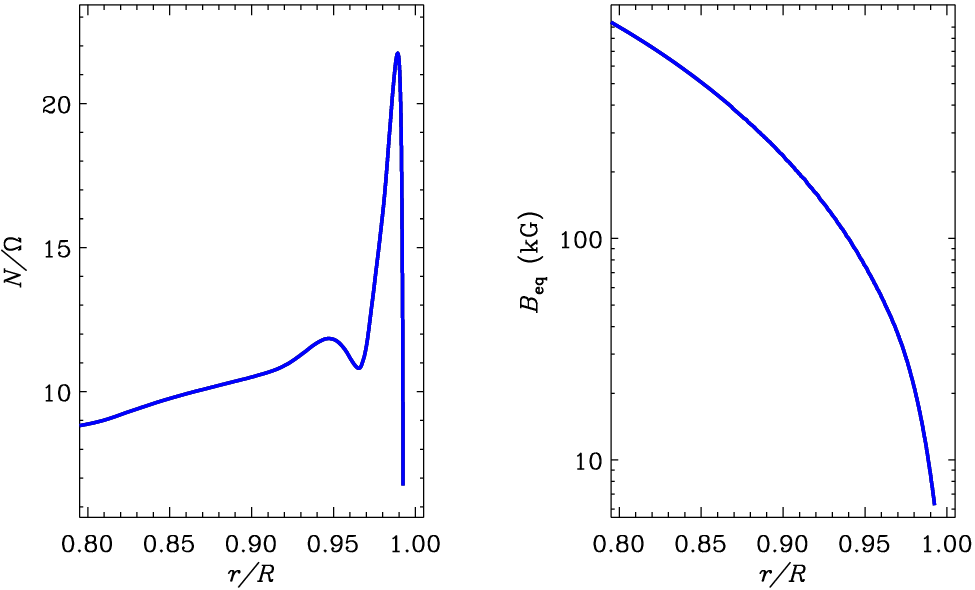}
\caption{The frequency ratio $N/\Omega$ (left panel) and the equipartition field strength (right panel) as a function of radius.}
\label{f1}
\end{figure}

Radial displacements in the stably stratified (entropy gradient $\partial S/\partial r > 0$) radiative zone are inhibited by buoyancy forces. Therefore, the radial scale of the perturbations is small with respect to the radius $r$. Our stability analysis is local in radius, but the spatial scale of the perturbations along the horizontal directions is unlimited. 

As usual, the linear stability analysis turns into an eigenvalue problem, i.e., searching the time dependence in the form $P,T \propto \exp (-\rm{i}\omega t)$. Eigenvalue
\begin{equation}
    \omega = \rm{i}\gamma + w
    \label{3}
\end{equation}
incorporates the growth rate $\gamma$ (decay rate if $\gamma < 0$) and the oscillation frequency $w$.

The magnetic field (\ref{1}) and stellar structure are assumed to be independent of longitude $\phi$. Therefore, perturbations with different azimuthal wave numbers $m$ ($\propto \exp({\rm i}m\phi$) develop independently, and their time and longitude dependence can be represented as $P,T \propto \exp[\gamma t + {\rm i}(m\phi - wt)]$. Any constant phase $m\phi - wt = const$ of a non-axisymmetric perturbation shifts in longitude with velocity $\dot{\phi} = w/m$. Perturbations with azimuthal wave number $m = 1$ have the largest growth rates of the Tayler instability \citep{Goossens_1981}. The calculations in this paper are limited to this case, where $w$ in Eq.\ref{3} is the drift velocity of the perturbation pattern in longitude. Simultaneously, the perturbation amplitude varies with the rate $\gamma$. 

\begin{figure}
\includegraphics[width=\linewidth]{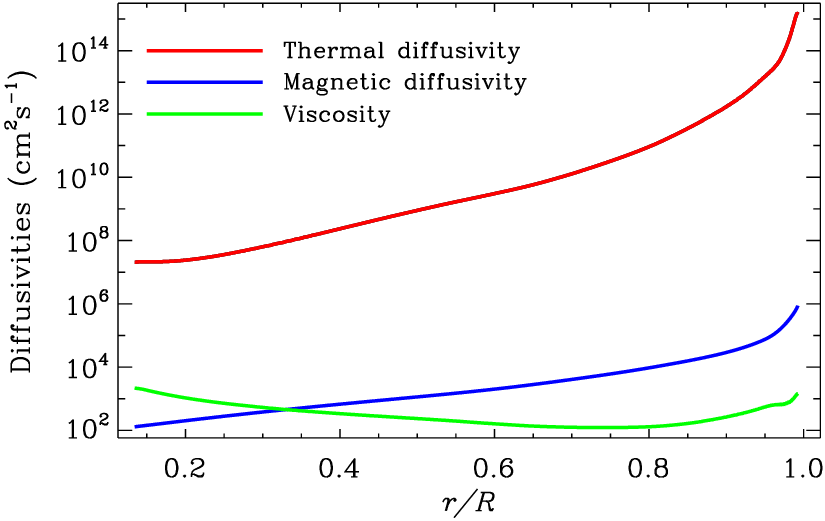}
\caption{Thermal diffusivity, magnetic diffusivity, and viscosity as functions of radius in the radiative zone of the star.}
\label{f2}
\end{figure}

The calculations are performed for a latitude-independent $\Omega_A$ in Eq.\ref{1}, i.e., for a background field symmetric about the equator.  

The control parameters of the model are the relative strength of the background field $\Omega_A/\Omega$ and the radial wavelength of the perturbation, 
\begin{equation}
     \hat{\lambda} = \frac{N}{\Omega kr} ,
     \label{4}
\end{equation}
where $N$ is the Brunt-V\"{a}is\"{a}l\"{a} frequency and $k$ is the radial wave number. The model also accounts for finite dissipation - viscosity ($\nu$), thermal diffusivity ($\chi$) and magnetic diffusivity ($\eta$) - via dimensionless parameters
\begin{equation}
    \epsilon_\nu = \frac{\nu N^2}{\Omega^3r^2},\ \ 
    \epsilon_\chi = \frac{\chi N^2}{\Omega^3r^2},\ \
    \epsilon_\eta = \frac{\eta N^2}{\Omega^3r^2}\,.
    \label{5}
\end{equation}

The stellar internal structure used to be known to find these parameters. It was obtained using the {\sl MESA} stellar structure and evolution model \citep{Paxton_2011}. Calculations for a star with mass $M = 3.2M_\odot$ and metallicity $Z = 0.015$ result in parameters close to those given in Table\,\ref{tab1}: $R/R_\odot = 2.325$ and $T_{eff} = 12585\,\rm{K}$ at the age $8.7\times10^7$ y. The parameters of the star for this age will be used in the calculations of the Tayler instability. The star has a central convective core with a radius of about $0.133R$, a thin ‘convective skin’\ with a thickness of $7.27\times 10^{-3}R$ on the surface, and an extended radiative zone between them, where the Tayler instability can grow. The parameters of the star necessary for the instability calculations are shown in Figs. \ref{f1} and \ref{f2}. Note that {\sl MESA} yeilds an increasing moment of inertia $I$ with characteristic time $I/\dot{I} \simeq 4\times10^8$~yr, which is too long to explain the observed deceleration of rotation.

The highest growth rate of the Tayler instability corresponds to $\hat{\lambda} \simeq 0.1$ in the Eq.\ref{4} \citep{Kitchatinov_2008a,Kitchatinov_2020}. Therefore, we take the stellar parameters for the instability calculations at a depth of $0.1R$ below the stellar surface, slightly larger than the wavelength $\lambda \simeq 2\pi r\Omega/(10N)$ of the most unstable perturbation. The frequency ratio $N/\Omega$ used in this estimation is shown in Fig. \ref{f1}. 

Figure \ref{f1} also illustrates the strength of the so-called equipartition field, when $\Omega_A = \Omega$ and the energy density of the toroidal field (Eq.\ref{1}) equals to the rotational kinetic energy. Comparison with the field strength in the Table\,\ref{tab1} yields that $\Omega_a/\Omega \approx 10^{-2}$ for 56 Ari.  

\begin{figure}
        \centering
        \begin{multicols}{1}
            {\includegraphics[width=2\linewidth]{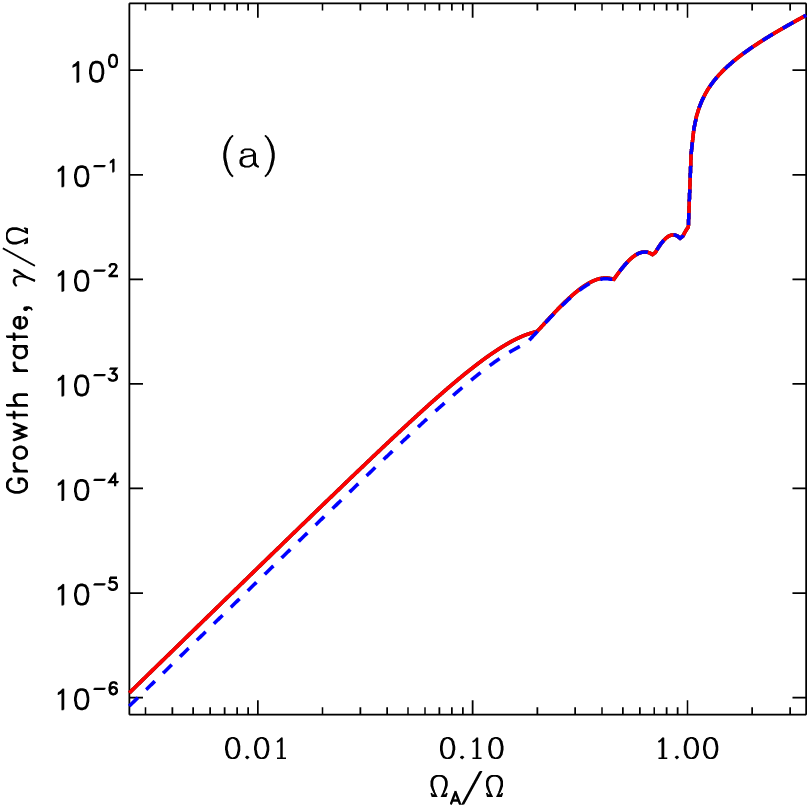}}\par 
        \end{multicols}
        \begin{multicols}{1}
            {\includegraphics[width=2\linewidth]{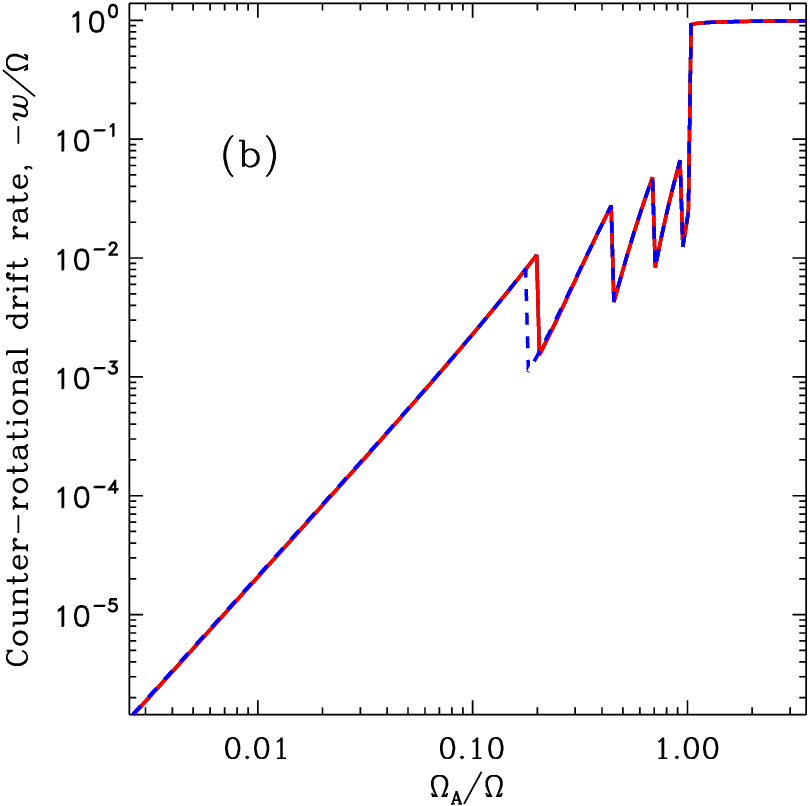}}\par
        \end{multicols}
    \caption{The growth (panel (a)) and drift (panel (b)) rates for the most rapidly growing modes of instability in the reference frame rotating with the star. The solid and dashed lines correspond to modes symmetric and antisymmetric about the equator, respectively.}
    \label{f3}
    \end{figure}


The dissipation coefficients used in the calculations are shown in Fig.\,\ref{f2}. The larger value of thermal diffusivity compared to magnetic one corresponds to the double-diffusive regime of magnetic buoyancy \citep{Parker_1979,Hughes_1995} of the background field. For the dissipation parameters (\ref{5}) we estimate $\epsilon_\nu\,\simeq\,1.5\times 10^{-14}$, $\epsilon_\chi\,\simeq\,8.9\times\,10^{-5}$, $\epsilon_\eta\,\simeq\,1.5\times\,10^{-12}$. Further calculations were performed for these values. Note that the large difference in the dissipation parameters (Eq.\ref{5}) led to difficulties in the numerical calculations for the equatorially asymmetric background field \citep{Kitchatinov_2020}. There are no such difficulties for the equatorially symmetric field (Eq.\ref{1}).

\begin{figure}
\includegraphics[width=\linewidth]{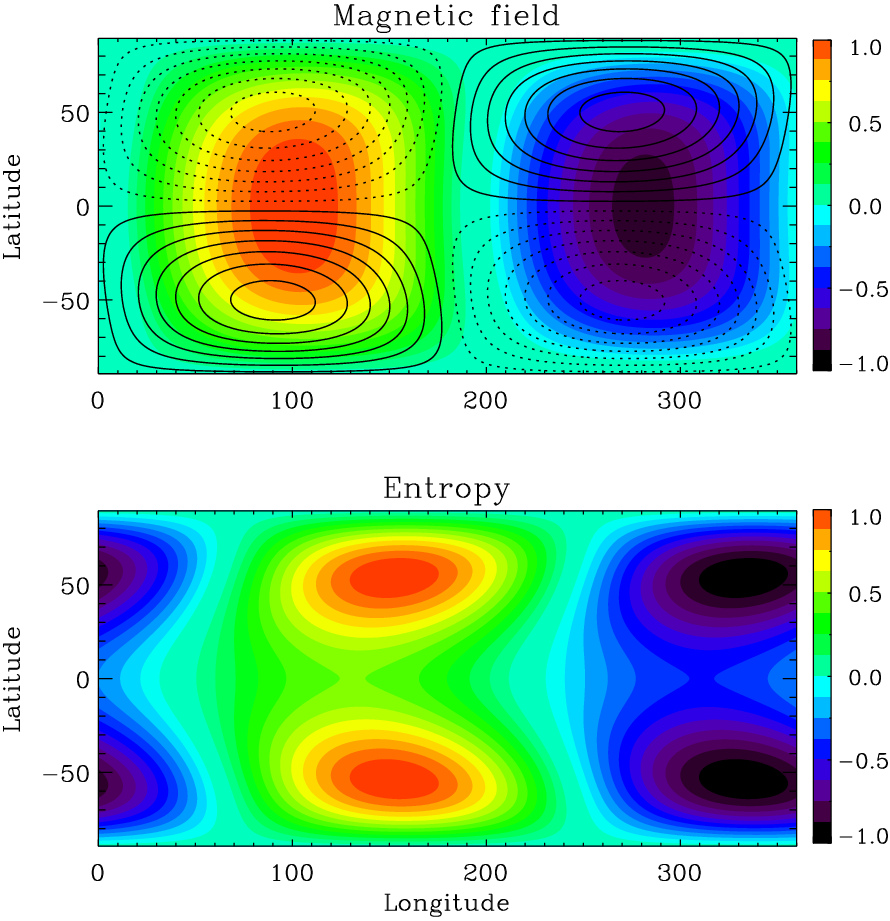}
\caption{Structure of the most rapidly growing mode of the instability for $\Omega_A/\Omega=0.01$. Upper panel: structure of the magnetic field. The colour scale shows the normalised strength of the radial field. The solid (dashed) lines show the force lines of toroidal field perturbations with clockwise (counterclockwise) circulation. Bottom panel: entropy (temperature) perturbations. The colour scales are graduated in arbitrary units.}
\label{f4}
\end{figure}

\section{Results and Discussion}

In our model, we imply a direct link between magnetic and chemical surface structures in the atmospheres of Ap/Bp stars. How quick can be the "response"\ of the spot structure to changes in the magnetic field driven by the Tayler instability? In the considered temperature range, silicon is represented primarily by the first ionisation stage in the line formation region: the ratio of concentrations $\log(\ion{Si}{I}/\ion{Si}{II})\sim10^{-3}-10^{-4}$ at $\log\tau\lesssim0.1$ \citep{Michaud_1981}. The collision frequency of silicon ions with protons $t_c$ can be estimated for the line formation region in the stellar atmosphere with $T_{eff}=12800$ K as $\log t_c\sim0.7$ c$^{-1}$ assuming a Maxwell velocity distribution. In turn, the gyrofrequency of the silicon ion \ion{Si}{II} in a magnetic field with strength $B\sim10^3$ G turns out to be of the order of $\omega_{ci}\sim10^5$ rad/s. Thus, with the $\omega_{ci}\gg t_c$, the magnetic field is frozen in the ionised component and controls its motion. Variations in the geometry and angular velocity of the magnetic structures as a result of instability will simultaneously lead to corresponding changes in the surface silicon spots, in turn affecting the photometric period.

Fig.\,\ref{f3} shows the growth rate $\gamma$ and drift rate $w$ of unstable modes as a function of the background magnetic field amplitude. There are, as usual \citep{Kitchatinov_2008a}, two types of modes, symmetric and antisymmetric with respect to the equator. Furthermore, for each symmetry type there is a discrete set of eigenmodes with different growth and drift rates, characterised by their own spatial structure. Fig.\,\ref{f3} shows the results for the “dominant”\ instability modes with the largest growth rates. The solid (dashed) lines correspond to equatorially symmetric (antisymmetric) modes.

Oscillations in the range $0.2 \lesssim \Omega_A/\Omega \lesssim 1$ arise due to switch of the dominant mode with the largest growth rate. At the same time, the $\gamma$ value remains a continuous function of the field strength. However, the drift velocity suffers a discontinuity. The sawtooth-shaped section of the curve in the right panel of the figure is related to these developments. In the region $\Omega_A > 0.2\Omega$ in Fig.\,\ref{f3}, the results for both types of equatorial symmetry almost coincide. Because with field amplification, the eigenmodes of the instability concentrate at higher latitudes \citep{Kitchatinov_2008b,Kitchatinov_2020} and appear to be no longer "sensitive"\, to equator. However, the equatorial-symmetric mode dominates in considered region of the relatively weak fields with $\Omega_A/\Omega \approx 0.01$. Its structure is shown in Fig.\,\ref{f4}. In linear analysis of stability, the amplitude of the perturbations remains uncertain. Therefore, the eigenmode of instability in Fig.\,\ref{f4} is represented in arbitrary units.
The figure reveals that the magnetic poles lie in the equatorial plane, i.e., the radial component of the field is a dipole with $\beta\approx90^\circ$, that is fully consistent with the observations of 56 Ari.

As one can see from Fig.\,\ref{f3}, the growth rate for the background field $\Omega_A/\Omega\approx0.01$ is $\gamma\approx10^{-5}\Omega$. Taking $\Omega=2\pi/\bar P_{rot}\approx$8.6 d$^{-1}$, results in instability growth time of $\sim30$ yr in the case of 56 Ari. The drift velocity in Fig.\,\ref{f3} has a negative sign. This means that in the reference frame rotating with the star, the instability pattern moves opposite to the direction of rotation. In the reference frame of the external observer its angular velocity is $\Omega_{obs} = \Omega + w$. The limiting cases $w/\Omega=0$ and $w/\Omega = -1$ correspond to the corotation of the instability picture with the star and its rest for the external observer\footnote{the latter case in the strong-field regime was previously proposed to explain the phenomenon of super-slowly rotating (SSRAp) Ap stars \citep{Kitchatinov_2020}}. For a weak background field $\Omega_A/\Omega\approx0.01$ the drift velocity is $w/\Omega\approx -10^{-5}$, which is close to corotation with the star. It should be emphasised that the instability calculations give the value of the drift velocity for the given value of $\Omega_A/\Omega$, but not its temporal variation, which is necessary to explain the changes in the photometric period. 

In the framework of discussed model, temporal changes in $w$ are possible in the following cases:
\begin{itemize}
 \item The development of instability occurs during the period of observations
 \item The changes in the drift velocity are due to the dynamic evolution of the background field
\end{itemize}

The abovementioned estimation of the instability growth timescale ($\sim30$ yr) show that it is possible to assume its development over the historical period of 56 Ari observations, although this scenario seems to be unlikely. Although magnetic Ap/Bp stars occupy the entire range of ages $t\sim10^6-10^9$ yr - prior to and on the MS, the instability growth time is many orders of magnitude shorter, as well as the probability of accidentally find a star in this phase. Let us estimate it as follows. Let $t_{inst}\sim10^2$ yr be the upper limit for the development of the Tayler instability in a star. $N_{inst}$ is the number of Ap/Bp stars undergoing instability growth at the present time. $t_{MS}\sim10^9$ is the lifetime of an A0-class star at the MS, $N_{MS}$ is the number of Ap/Bp stars at the MS. Then, from the proportion $N_{inst}/N_{MS}=t_{inst}/t_{MS}$, we obtain an ratio of the number of Ap/Bp stars at the stage of the development of the Tayler instability to the total number of Ap/Bp stars on the MS as $N_{inst}/N_{MS}\approx10^{-7}$. At the same time, in the local sample, the number of Ap/Bp stars with variable periods of order 10 \citep{Pyper_2021} that is much larger than this estimate and amounts to $\sim0.3$\% of their total number \citep{Renson_2009}). 

Another possibility for the variation of the drift rate $w$ is the variation of the background field producing the instability. In the considered region of Fig.\,\ref{f3}, in the sub-Alfven region, $w \propto \Omega_A^2$, therefore the drift velocity of the instability pattern can increase with relatively small variations of the background field. Adopting the rate of the period changes in 56 Ari as $\Delta P\approx4$ s/100 yr, we estimate the change $\Delta w/\Omega\sim10^{-4}$. From the graph in Fig. \ref{f3} this corresponds to changes in the magnetic field $\Omega_A/\Omega$ of order of the factor 2. Such changes are not impossible, although discussion of their causes is somewhat speculative. Let us point out two principal possibilities. Thus, one can assume the field strengthening under the action of the so-called "Tayler-Spruit dynamo"\, \citep{Spruit_2002}. Within this scenario, the dynamo cycle is induced by the generation of a toroidal magnetic field by the differential rotation ($\Omega$-effect), this field in turn is subject for the Tayler instability. Within the closed dynamo loop, the field amplification occurs, which in turn leads to a change in the drift velocity of the non-axisymmetric mode of the Tayler instability.

The other possibility for the fluctuations in field strength, leading to the same effect for the drift velocity, is the double-diffusive mechanism of uplift to the overlying layers of magnetised volumes under the combined action of magnetic buoyancy and thermal diffusion \citep{Parker_1979,Hughes_1995}.

The important source of uncertainty in our approach is the lack of any observationally confirmed information on the distribution and characteristics of the background magnetic field of the star. The equatorially symmetric geometry of the background field (Eq.\ref{1}) adopted in our calculations allows to reproduce the observed surface geometry of the field. It is consistent with the concept of Ap stars formation from the stellar mergers \citep{Bogomazov_2009,Tutukov_2010,Schneider_2019}: the dynamo in accretion disc produces a large-scale field symmetric with respect to its central plane of the disc \citep{Rudiger_1995,Rekowski_2000}. This field settles with the disc matter on the accreting star and produces an equatorially symmetric magnetic configuration. However, the details of the latitudinal distribution of the field and its small-scale structure remain unknown. While the latitudinal distribution primarily affects the quantitative characteristics of the instability, but not its global pattern, the possible mechanism of generation of field strength fluctuations depends on the fine structure. The amplification of the large-scale background field due to the double-diffusive instability mechanism is possible if field consists of the small-scale magnetic tubes. 

It is worth noting that any models of the magnetic fields of stars of early spectral classes, including the competing group of models explaining changes in the periods of Ap/Bp stars by Alfvén waves, face similar problems. Thus, in the \citep{Takahashi_2024} models, the internal structure of the stellar magnetic field, which provides frequencies close to the observed ones, appears to be model-dependent and, as the authors themselves note, there is also no clarity about the initial mechanism of excitation of Alfvén waves. The possibility of acquiring new information about the internal fields of stars seems to be related to the development of astroseismology methods and their combination with spectropolarimetry \citep{Prat_2019,Mathis_2021}. However, currently we have to compare the predictions of the models with observations of the surface magnetic fields and it indirect manifestations, such as the inhomogeneous horizontal distribution of chemical elements and the rotational modulation of the radiation flux in Ap/Bp stars.
In turn, the photometric observations, although clearly indicate the period variations in 56 Ari and other Ap/Bp stars from the discussed subgroup, but admit different interpretations of the character of these variations: linear and discrete, or oscillating. Various mechanisms leading to these changes can be proposed: Tayler instability and Alfvén waves, respectively. Note that in a certain range of field strengths, the Tayler instability leads to a quasi-periodic change of the drift rate due to the change of the dominant instability mode. Potentially, this character of $w$ changes could also explain the discrete period changes observed in CU Vir.

In summary, magneto-rotational effects, including the Tayler instability, are currently the most promising group of mechanisms responsible for changes in the periods of Ap/Bp stars. This means that the magnetic fields of Ap/Bp stars are apparently not as stable as previously thought and are subject to MHD instabilities with observable manifestations. For a more definite choice of the physical mechanism, it is necessary to continue the series of photometric observations and to improve their accuracy.

\section{Conclusions}
The star 56 Ari belongs to a small subgroup of magnetic Ap/Bp stars showing changing photometric periods, which have no explanation in the framework of the current view on the evolution of angular momentum and mechanisms of its loss in single stars. These period changes are larger than the expected rate of angular momentum loss and their unambiguous physical interpretation is currently lacking. We suggest that the observed periods changes could be due to the azimuthal drift of surface magnetic structures and associated chemical spots caused by the Tayler instability of the magnetic field in the radiative envelopes of Ap/Bp stars. The MHD modelling of instability performed for the case of 56 Ari yielded a surface field geometry with a large inclination of the dipole axis to the rotational one, consistent with the observations and indirectly indicating a possible scenario of the field generation as a result of stellar merging. The growth rate of the instability appears to be much shorter than the lifetime of the star on the MS, and the longitudinal drift of its most rapidly growing mode is directed opposite to the stellar rotation, that corresponds to the observational effect of increase of the photometric period. The calculated drift rates make it possible to explain the rate of the period changes in 56 Ari by the rigid shift of magnetic and spot structures, assuming small variations in the background field.

The choice of a specific mechanism causing the observed changes in the periods of 56 Ari and other similar Ap/Bp stars demands both more accurate and extended observational series and direct information on the distribution of the magnetic field in the stellar interior. In the present paper we show the principal possibility of the explanation of this phenomenon by the Tayler instability of magnetic field.

\section*{Acknowledgements}

This work was financially supported by the Ministry of Science and Higher Education of the Russian Federation.\\




\bibliographystyle{mnras}
\bibliography{references.bib} 


\bsp	
\label{lastpage}
\end{document}